\documentclass{elsart}
\usepackage{amsmath}
\usepackage[dvips]{graphicx}

\begin{document}

\newcommand{\BE}{\begin{eqnarray}}
\newcommand{\EE}{\end{eqnarray}}
\newcommand{\BEn}{\begin{eqnarray*}}
\newcommand{\EEn}{\end{eqnarray*}}
\newcommand{\barr}{\begin{array}} 
\newcommand{\earr}{\end{array}}
\newcommand{\bit}{\begin{itemize}}      
\newcommand{\eit}{\end{itemize}}
\newcommand{\bfl}{\begin{flusleft}}
\newcommand{\efl}{\end{flusleft}}
\newcommand{\bfr}{\begin{flushright}}
\newcommand{\efr}{\end{flushright}}

\newcommand{\bc}{\begin{center}}
\newcommand{\ec}{\end{center}}

\newcommand{\ben}{\begin{enumerate}}    
\newcommand{\een}{\end{enumerate}}

\newcommand{\cl}{\centerline}
\newcommand{\ul}{\underline}
\newcommand{\nl}{\newline}


\newcommand{\impl}{\Longrightarrow}
\newcommand{\eps}{\varepsilon}
\newcommand{\de}{\partial}


\newcommand{\xb}{{\bf x}}
\newcommand{\rb}{{\bf R}}
\newcommand{\kb}{{\bf k}}
\newcommand{\qb}{{\bf q}}
\newcommand{\eb}{{\bf E}}

\begin{frontmatter}
\title{Synchronization of non-chaotic dynamical systems}
\author[dma,INFMFI]{Franco Bagnoli\corauthref{cor1}}
\ead{bagnoli@dma.unifi.it}
\author[sissa,INFMTS]{Fabio Cecconi}
\ead{cecconi@sissa.it}
\corauth[cor1]{to whom correspondence should be addressed}
\address[dma]{Dipartimento di Matematica Applicata, Universit\`a di
         Firenze, Via S.~Marta, 3 I-50139 Firenze}
\address[sissa]{International School for Advanced Studies (SISSA/ISAS)
         via Beirut 2-4, I-34014 Trieste}
\address[INFMFI]{INFM Unit\`a di Ricerca di Firenze}
\address[INFMTS]{INFM Unit\`a di Ricerca di Trieste (Sissa)}

\begin{abstract}
A synchronization mechanism driven by annealed noise is studied for 
two replicas of a coupled-map lattice which exhibits \emph{stable 
chaos} (SC), i.e.\ irregular behavior despite a negative Lyapunov spectrum.  
We show that the observed synchronization transition, on changing
the strength of the stochastic coupling between replicas, 
belongs to the directed percolation universality class. This
result is consistent with the behavior of \emph{chaotic} deterministic
cellular
automata (DCA), supporting the equivalence Ansatz between SC models and
DCA.
The coupling threshold above which the two system replicas synchronize 
is strictly related to the propagation velocity of perturbations in the system. 
\end{abstract}
\end{frontmatter}

\section{Introduction}

The occurrence of disordered patterns and their propagation in the presence 
of a negative Lyapunov spectrum have been often observed in
spatiotemporal systems~\cite{CK88,PLOK,mappets,BP,Lai95}.
One can classify this kind of irregular behavior into two general 
groups with basically different features: {\em transient chaos} 
and {\em stable chaos}.

Transient chaos (TC) is a truly chaotic regime with finite lifetime, 
characterized by the coexistence in the phase space of stable attractors 
and chaotic non attracting sets - named chaotic saddles or 
repellers~\cite{Tel}.
The system, starting from a generic configuration, typically exhibits 
irregular behavior until it collapses abruptly onto a nonchaotic attractor. 

Stable chaos (SC) constitutes a different kind of transient
irregular behavior \cite{CK88,PLOK} which cannot be ascribed to the
presence of chaotic saddles and therefore to divergence of nearby
trajectories.
In SC systems, moreover, the time spent in transient regimes may scale
exponentially with the system size (supertransients~\cite{CK88,PLOK}),
and the final stable attractor is practically never reached for large
enough systems.
One is thus allowed to assume that such transients may be of substantial 
experimental interest and become the only physically relevant states in the 
thermodynamic limit. 
While TC remains associated to information production, i.e.\ to the
response of the system to infinitesimal disturbances, SC is mainly related
to propagation and mixing of information. 
In other words, SC systems are sensitive only to  
perturbations of finite amplitude~\cite{PT}, 
while they respond to infinitesimal disturbances in a way 
similar to stable systems. 
Such a key feature makes meaningless any characterization of the SC
complexity by means of Lyapunov theory.

In this paper we focus on SC behavior whose origin has not yet
found a convincing explanation, even though, it has been observed in
several spatially-extended models such as coupled map
lattices~\cite{CK88,PLOK,mappets} and oscillators~\cite{BP}.
To provide an adequate description of SC systems we invoke their strict
similarity with discrete models such as deterministic cellular automata
(DCA).
In fact, according to the conjecture that ``SC systems represent 
a continuous generalization of deterministic cellular automata''
\cite{PLOK}, we can argue that, what is known about DCA could be
automatically translated into the language of SC.
Although a general mapping of SC- onto DCA-models is still missing,
the conjecture is supported by the fact that, also in finite size DCA,
limit cycles and fixed points are the only allowed attractors, since the
number of possible configurations is finite.
Moreover in some DCA, the transient dynamics towards the final attractor
may exhibits a long living irregular behavior with lifetimes that typically
grow exponentially with the system size. In this case it is practically
impossible to find any recurrence (Poincar\'e cycles) for large systems.
According to Wolfram classification~\cite{Wol}, these DCA form the
third (``chaotic'') class and they share several properties with 
continuous SC systems.
 
The emergence of this ``chaoticity'' in DCA dynamics, is
effectively illustrated by the damage spreading
analysis~\cite{Damage1,Damage2}, which measures the sensitivity to initial
conditions and for this reason is considered as the natural
extension of the Lyapunov technique to discrete systems.
In this method, indeed, one monitors the behavior of the distance between 
two replicas of the system evolving from slightly different initial
conditions.
The dynamics is considered unstable and the DCA is said chaotic, whenever
a small initial difference between replicas spreads through the whole system. 
On the contrary, if the initial difference eventually freezes or disappears, 
the DCA is considered non chaotic.  

A similar scenario holds for systems exhibiting SC~\cite{mappets}, 
where a transitions between regular and irregular dynamics may 
occur upon changing a control parameter (e.g. the spatial coupling 
between sites); the irregular dynamics is often associated to 
spreading of damages.

In this work we show that another characterization of the SC behavior 
can be achieved through a suitable synchronization method, which has 
been successfully employed in Ref.~\cite{BR} to classify the chaotic
properties of DCA. 
In our opinion this method may be used to complement the common damage 
spreading analysis. 
The basic idea consists of measuring the minimal ``strength'' of the 
coupling between a replica (slave) and the original system (master) required  
to achieve their perfect synchronization.  
The master-slave interaction is a stochastic and spatially-extended version of
the Pecora-Carroll synchronization mechanism~\cite{Pecora}.
Unlike the damage spreading method, where replicas are independent, this 
coupling scheme implies that the evolution of slave is driven by the master. 
Practically each time step is composed by two phases: in the first one,   
the master and slave system evolve freely with the same
evolution equation, and    
then a fraction $p$ of the degrees of freedom in the slave-system is enforced
to take the value of the corresponding degrees of freedom in the
master.
Synchronization of spatially extended systems has been usually studied
with symmetrical interactions~\cite{Amos,Zan,Morel,Bocca,Grass99}. Our
asymmetric scheme, instead, allows probing the dynamical properties of the
master system. 
Through a gradual increase of $p$ from $0$ to $1$, the dynamics of the slave 
system tends to synchronize to that of master, and at a threshold $p^*$ a 
synchronization transition occurs: the pinching synchronization transition 
(PST).
The threshold $p^*$ above which the replicas synchronize, is an indicator 
of the chaotic behavior of the unperturbed system (master). 
Indeed, a large value of $p^*$ implies a large fraction of sites to be 
pinched in order to achieve the synchronization, indicating that the
dynamics of the replicas is rather irregular and difficult to control.

In DCA, the PST belongs to the directed percolation (DP)
universality class~\cite{BR,Grass99}, while in fully chaotic continuous systems 
(e.g. coupled map lattices) the synchronization is never 
perfect for finite times, nor it is equivalent to an absorbing 
state~\cite{Grass99}. This generally implies non-DP scaling
exponents~\cite{Grass95,Piko,Grin}.
Here, we find that the PST is well defined for a model showing typical
SC behavior originally introduced in Ref.~\cite{PLOK}. This PST is found
to belong to the DP universality class, in agreement to what happens for
DCAs. The result further supports the conjecture that SC class contains DCAs. 

The sketch of this paper is the following.
In the next section, we study a simple coupled-map lattice which is
particularly suitable for discussing the distinction between SC and TC behaviour.
The model, in fact, displays a TC regime before falling into DCA
dynamics with the typical properties of a discrete SC regime.
In Section~III, we describe the SC model of Ref.~\cite{PLOK},
whose dynamics never reduces to a DCA, and the pinching synchronization
technique applied to it. For the latter model we obtain the PST phase
diagram (Sec.~IV), by measuring the synchronization threshold as a
function of the coupling strength between sites.
Such a phase diagram agrees remarkably with that already obtained through
damage spreading analysis in Ref.~\cite{mappets}. We provide an argument to
explain this consistency.
Finally, conclusions and remarks are reported in the last section.

\section{Transient and stable chaos} 
\label{sec:pedagogical}

It is instructive to discuss the main differences between SC and TC regimes
with the aid of a simple spatiotemporal model in which both of them occur.

Let us consider the one-dimensional coupled-map lattice (CML),
i.e. an array of state variables $\{x_1,...,x_L\}$ in the interval 
$[0,1]$ subject to the discrete-time evolution rule
\begin{equation}
x_i(t+1) = g_{\eps}(x_{i-1}(t),x_{i}(t),x_{i+1}(t)). 
\label{eq:cml}
\end{equation}
Each site-variable $x_i$ interacts diffusively with its 
nearest neighbors 
\begin{equation}
g_{\eps}(u,v,z) = (1 - 2 \eps) f(v) + \eps \big[ f(u) + f(z) \big].
\label{eq:stepcml}
\end{equation}
where $\eps$ sets the coupling strength among maps.

The local mapping has the form shown in Fig.~\ref{fig:stepmap}:
\begin{equation}
f(x) = \left\{\begin{aligned}
	0  & \qquad\text{for $0\le x < \alpha$,}  \\
\dfrac{x-\alpha}{1/3-2\alpha} & \qquad\text{for $\alpha\le x < 1/3-\alpha$,}\\
	1 & \qquad\text{for $1/3-\alpha \le x < 1/3+\alpha$,}\\
	1-\dfrac{x-1/3-\alpha}{1/3-2\alpha} & \qquad\text{for $1/3+\alpha\le x <
	2/3-\alpha$,}\\
	0 & \qquad\text{for $2/3-\alpha\le x < 2/3+\alpha$,}\\
	\dfrac{x-2/3-\alpha}{1/3-2\alpha} & 
       \qquad \text{for $2/3+\alpha\le x < 1-\alpha$,}\\
	1 & \qquad\text{for $1-\alpha\le x < 1$.}
	\end{aligned}\right.
\label{eq:stepmap}
\end{equation}
In this example, we always use the democratic coupling $\eps=1/3$,
and we shall neglect to indicate the $\eps$ dependence.
A typical grayscale pattern generated by the evolution of the above CML is
shown in Fig.~\ref{fig:peda} (see the caption for the grayscale-code), 
for $\alpha=0.068$. Note that the continuous dynamics (gray)
is limited to restricted domains, while in the rest, the system has fallen
into a Boolean dynamics (typical of DCA), which is however far from being
trivial.

Systems showing SC and TC regimes are asymptotically stable, therefore 
the standard Lyapunov spectrum is not able to detect their transiently
irregular states.
In TC regimes, however, the existence of chaotic saddles in the phase 
space is revealed by the finite-time (or effective) Lyapunov exponent
\begin{equation}
\gamma(t) = \frac{1}{t}
\langle \log \|{\bf w}(t)\| \rangle,
\label{eq:lyap}
\end{equation}
where ${\bf w}(t) = \{w_1(t),...,w_L(t)\}$  indicates a generic
infinitesimal perturbation (i.e., a tangent vector) at time $t$,
which evolves following the linearization of
Eqs.~(\ref{eq:cml}--\ref{eq:stepmap}).
The average in expression~(\ref{eq:lyap}) is taken over the ensemble of
trajectories which have not yet left a certain neighborhood of the
saddle at time $t$ \cite{Ott}. The indicator $\gamma(t)$ is expected to
fluctuate around a positive value during the transient~\cite{Lai95} 
and switches to negative values after the transition to the stable 
attractor occurs. 
In SC, however, even the finite-time Lyapunov exponent does not provide
much information, as it becomes negative already in the transients.
This indicates that the source of the SC behaviour cannot rely on the instability associated
to repelling sets. Although, the existence of repellers may not be excluded 
{\it a priori}, certainly their role is not observable in SC.

We now see how the above considerations apply to our toy-model.  
First, we discuss the case of infinite slope map, (i.e.\ $\alpha = 1/6$,
full line in Fig.~\ref{fig:stepmap}) corresponding to a pure SC regime.
In fact, after one time step, each  configuration of the lattice
reduces to a sequence of ``0'' and ``1'' (Boolean configuration).
The system evolution remains, however, irregular since,
when $\alpha=1/6$, the model is equivalent to a
DCA which follows the {\em rule 150}
\[
 g(u,v,z) = u + v + z - 2(uv+uz+vz) + 4uvz
\]
with $u$,$v$ and $z$ Boolean variables.
This dynamics is known to generate highly irregular patterns~\cite{Wol}.
On the other hand, this irregular behaviour cannot be associated to 
either chaotic saddles or local fluctuations of Lyapunov exponent,
the latter being $-\infty$ due to the specific form of the map. 
Despite Lyapunov analysis ensures that this system is totally 
insensitive to infinitesimal perturbations, finite perturbations of
amplitude greater than $1/6$ give rise to a ``defect'' which propagates
through the lattice.
Indeed, it can be easily shown that a defect also evolves with the
chaotic {\em rule 150}, because the rule is additive modulo two.

A slight tilting of the vertical edges of the map, ($0<\alpha<1/6$, 
dotted and dashed lines in Fig.~\ref{fig:stepmap}) 
introduces some expanding regions in the phase space.
Accordingly, one obtains a typical TC behaviour, due to the competition
between stable and unstable effects, which decays into the above mentioned
SC regime (see Fig.~\ref{fig:peda}).
Figure~\ref{fig:multipl} shows a typical time fluctuation of the local
expansion rate (or local multiplier)
$\mu(t) = \|{\bf w}(t)\|/\|{\bf w}(t-1)\|$, for $\alpha=0.052$,
${\bf w}$ being a tangent vector as in Eq.~({eq:lyap}).   
The irregularity of the signal is a clear indication of the chaotic-like
behavior of the system.
The simulation is stopped at time $T_r$ when ${\bf w}(t)$ becomes
exactly $\bf 0$, i.e.,\ when the system settles down into
a Boolean configuration (SC state) which is Lyapunov-stable by definition.  
Therefore, $T_r$ provides an estimate of the time spent in the TC regime.

For all those trajectories which have not yet entered binary configurations
at time $t$, $\gamma(t)$ remains positive, as seen in Fig.~\ref{fig:hist},
where the distribution of $\gamma(t)$ is shown for a set of $2000$
trajectories starting from arbitrary initial conditions. 

The lifetime $T_r$ of the TC regimes preceding the SC behavior depends 
on $\alpha$. $T_r$ becomes shorter as $\alpha$ approaches $1/6$, value at 
which it vanishes, because the system behaves as a true DCA after just one
time step.
In the limit $\alpha \to 0$, instead, the flat regions of the map disappear 
(the slope being equal to $3$) and TC regime becomes persistent and  
degenerates into fully developed chaos (FDC).

A more detailed analysis of the behavior of the model of 
Eqs.~(\ref{eq:cml}--\ref{eq:stepmap}) 
upon changing the control parameter $\alpha$ will be presented elsewhere. 
Here such a qualitative discussion has the only aim to highlight the key 
differences between SC and TC dynamics, both present in this toy model.

In the example discussed so far, the SC regime occurs only as a DCA
behavior, since, as soon as the system falls into a 
Boolean configuration, it evolves as a genuine ``chaotic'' DCA. 
In the following section we discuss a continuous SC model whose behavior 
never reduces to DCA dynamics. In particular we study the 
synchronization properties of two replicas of such a system.

\section{The model and the synchronization dynamics} 

The dynamical system considered now, is the one-dimensional CML of
Eqs.~(\ref{eq:cml},\ref{eq:stepcml})
with the coupling constant 
$\eps \in [0,1/2]$ and periodic boundary 
conditions over a length $L$ (system size). 
The local mapping has the form
\begin{equation}
f(x) = \left\{\begin{aligned}
       bx         &\qquad \text{if $0 < x < 1/b$} \\
     a + c(x-1/b) &\qquad \text{if $ 1/b < x < 1$}\\
		\end{aligned} \right.
\label{eq:map}
\end{equation}
as shown in Fig.~\ref{fig:figmappet}.

We use here the parameter values ~$(a=0.07,\,b=2.70,\,c=0.10)$~ of
Ref.~\cite{mappets} for which the map of Eq.~(\ref{eq:map}) is attracted into a
stable period-3 orbit. An example of the space-time evolution of the CML is
shown in Fig.~\ref{fig:mappet}, where the presence of propagating structures
similar to those in Fig.~\ref{fig:peda} is observed, despite here the
system dynamics never relaxes onto a pure DCA state. 

In Ref.~\cite{mappets} an $\eps$-dependent dynamical phase transition 
between periodic and chaotic regimes of this system has been 
carefully investigated by damage spreading analysis.  
The periodic (chaotic) phase is characterized 
by the absence (presence) of damage propagation which is found to
behave linearly
\begin{equation}
S(t) = S(0) + 2V_F t
\label{eq:damage}
\end{equation}
being $S$ the linear size of the region affected by the
damage. The factor $2$ is a consequence of the symmetric coupling scheme
which requires the left and right damage-front to progress at the same
velocity $V_F$, but in opposite directions.
The damage spreading velocity $V_F$ can be considered a good
indicator for this transition, because it vanishes in periodic phases.
It turns out from Ref.~\cite{mappets},
that for $\eps < \eps_c^{(1)} \simeq 0.3$ only periodic phases are observed, 
for $\eps> \eps_c^{(2)} \simeq 0.3005$ only chaotic states exist.
In the intermediate region 
$\eps \in [\eps_c^{(1)},\eps_c^{(2)}]$, periodic and chaotic behaviors
alternate in an apparently irregular manner (fuzzy region).

We study the effects of the pinching synchronization 
on this model, and its interplay with the above described transition. 

The master system follows the dynamics of Eq.~(\ref{eq:cml}), while the slave system 
evolves as 
\begin{equation}
y_i(t+1) = [1-r_i(t)]g_{\eps}(y_{i-1}(t),y_{i}(t),y_{i+1}(t)) + 
r_i(t)g_{\eps}(x_{i-1}(t),x_{i}(t),x_{i+1}(t)) 
\label{eq:slave}
\end{equation}
where $r_i(t)$ is a Boolean random variable
\begin{equation*}
r_i(t) = \left\{ \begin{aligned}
	       1 &\quad \text{with probability  $p$},    \\
         0 &\quad  \text{otherwise}.
			\end{aligned}\right.
\end{equation*}

Practically, at each time step, a fraction $p$ of site variables in the slave
system is set equal to the  corresponding variables of the master 
(see. Fig.~\ref{fig:pinch}). 
In the limit case $p=0$, the slave system evolves independently of the 
master, while for $p=1$, its evolution coincides with the master one.   

The synchronization order-parameter is the asymptotic value of the 
topological distance $\rho$ between master and slave systems 
(i.e.\ the fraction of non synchronized sites)
\begin{equation}
\rho(t,p) = \lim_{L \rightarrow \infty} \frac{1}{L} \sum_{i=1}^{L} \Theta(|x_i(t)-y_i(t)|)
\label{eq:distop}
\end{equation}  
where $\Theta(s)$ is the unitary step-function.
We denote by $p^*$ the synchronization
threshold, such that $\rho(\infty,p<p^*) >0$ and $\rho(\infty,p<p^*)=0$. 
This synchronization mechanism defines an associated directed
site-percolation problem in $d=1+1$ dimension, where a site of 
coordinate $(i,t)$ is ``wet'' if 
$r_i(t)=0$ and it is connected to at least one neighboring wet site 
at time $t-1$.
At $t=0$ all sites are assumed to be wet. We denote by  $p_c$ the critical 
threshold for which a  cluster of wet sites percolates along the time
direction. The master and the slave systems can stay different
only on the cluster of wet sites.

For chaotic CMLs two 
synchronization scenarios are possible, called {\em weak} 
and {\em strong} chaos respectively~\cite{BBP}. A system is said strongly
chaotic if it does not synchronize even on the critical wet cluster 
(i.e.\ for $p^*=p_c$) and therefore the distance $\rho$ of Eq.~(\ref{eq:distop}) 
always exhibits DP scaling. Alternatively one can say that for strongly chaotic 
systems, the active and the wet clusters are essentially the same for every 
value of $p$.

Instead, for weakly chaotic 
systems, the synchronization threshold $p^*$ is always located below $p_c$ and 
the transition is discontinuous (first-order like). 
For $p<p^*$  again the active and 
wet clusters coincide, whereas for $p>p^*$ the active cluster disappears, but 
the wet cluster still percolates. Such a behavior is due to the exponential
vanishing of the difference field ${\bf h}(t)=\{x_i(t)-y_i(t)\}_{i=1}^L$, 
even though local fluctuations of ${\bf h}$ can sporadically appear.
       
However, for chaotic systems, the synchronized state is not robust
with respect to an infinitesimal perturbation in the absence of the
synchronization mechanism, i.e.\ it is not a proper absorbing state. 
Conversely, DCAs \cite{BR} always synchronize at $p^*$ below  $p_c$ 
(i.e.\ synchronization occurs in presence of the percolating wet cluster) but
$\rho(t,p)$ still exhibits DP scaling. This is a straightforward 
consequence of the finiteness of the number of states in DCAs, which prevents 
fluctuations of $\mathbf{h}$ in the absorbing state.
The toy model of Section~II trivially follows this behavior, and the
synchronized state is insensitive with respect to sufficiently small
perturbations. 
We show in the following that this scenario holds also for continuous
SC systems. 

\section{Numerical results}
A first set of simulations has been performed for lattices of size $L = 3000$, 
with periodic boundary conditions.
We measured $\rho(t,p)$ at different times and the results have been
averaged over a $5000$ randomly chosen initial conditions. 
For each simulation a transient of $10^4$ time steps has been discarded in 
order to avoid initial bias and reach stationary states.  
A site $i$ is considered to be synchronized (and $y_i(t+1)$ is set
equal to $x_i(t+1)$) if $|y_i(t)-x_i(t)|<\tau$, where $\tau$ is a
sensibility threshold. In this way we can control the
effects of the finite precision of computer numbers. We checked 
that the results are
independent of $\tau$ (for small $\tau$), and we 
choose $\tau=10^{-8}$ for massive simulations.

A typical behavior of $\rho(t,p)$ near $p^*$ is shown in Fig.~\ref{fig:rho}.
The value of $p$ corresponding to the most straight curve at large $t$ 
represents the best approximation of  $p^*$, which can be estimated
with good accuracy. 
The straight dashed line indicates the DP scaling results.  

For those values of $\eps$ for which the above analysis provided a  
too uncertain result in the estimation of $p^*$, we have carried out  
single-site simulations for larger systems, obtaining a more accurate 
determination of $p^*$.
These simulations consist in preparing the slave system exactly 
synchronized to the master except for the central site, and in
measuring 
how non-synchronized sites propagate throughout the system, generating DP-like 
clusters. This method probes the stability of synchronized states with 
respect to minimal perturbations. 
 
As usual in this type of simulations, we measured
the survival probability $P(t)$ of the desynchronized states, 
the number of non synchronized sites $N(t)$, and  their second 
moment $R^2(t)$ with respect to the center of the lattice 
(generally called the gyration radius).
Near the synchronization threshold and in the long time limit, these 
magnitudes  are expected to scale as \cite{Torre,Vespo}  
\BE
N(t) \sim t^{\eta}, \quad P(t) \sim t^{-\delta}, \quad R^2(t) \sim t^z. 
\EE

The determination of the asymptotic value of such exponents (e.g.\
$\eta$) is performed by plotting the effective exponent
\[
\overline{\eta}(t) = \dfrac{\log(N(a t))}{\log(N(t))}
\]
versus $1/t$ for several values of $p$. Here $a$ indicates an arbitrary
scale factor and we always set $a=2$. 
In the limit $t\rightarrow \infty$
$\overline{\eta}$ converges to $\eta$ for $p=p^*$, and diverges 
for other values of $p$. 
In pure DP systems this method allows the simulation of effectively
infinite lattices, 
since the reference state, i.e. the absorbing one (usually 
made of "0"s) is unique and does not change in time. 
Conversely, in our case the absorbing state coincides with 
the synchronized state and this implies   
the detailed knowledge of the evolution of the master system. 
This circumstance imposes severe limitations on 
lattice sizes and performances of the method. 

In Figure~\ref{fig:eff} we report the behavior of the effective exponents
$\overline{\eta}$, $\overline{\delta}$ and $\overline{z}+\overline{\eta}$
versus $1/t$ for several values of $p$. The asymptotic values of the
exponents
($\eta=0.330(5)$, $\delta=0.13(2)$, 
$z=1.25(2)$)
are consistent with the best known DP ones 
($\eta=0.31368(4)$, $\delta=0.15947(3)$, 
$z=1.26523(3)$~\cite{Vespo}).  
 
This analysis, repeated for several values of $\eps$, indicates that 
the synchronization of SC systems reasonably belongs to the DP universality
class.
The very slow convergence of the system dynamics to the asymptotic state 
makes hard
to exclude that for some values of $\eps$ the DP character of the
transition is violated.
However, we believe that, owing to the stability of the 
model of Eqs.~(\ref{eq:cml},\ref{eq:stepcml},\ref{eq:stepmap}),
small local disturbances are re-absorbed at exponential rate, and cannot   
generate desynchronization effects. In other words, the synchronized state is
absorbing with respect to small fluctuations.
This is also consistent with the
fact that the simulation results are 
independent of the precision threshold $\tau$.
Therefore, the local stability guarantees that, for what concerns 
synchronization properties, SC systems behave mainly like 
discrete ones and differently from continuous chaotic systems.
This further supports the ``{\em equivalence Ansatz}'' between SC and
DCA. 
Indeed, preliminary simulations show that the DP scenario holds even if the
local pinching is performed only up to a small difference $\Delta$. 
This remark suggests the possibility of defining a
finite-size maximum Lyapunov exponent, in a way similar 
to Ref.~\cite{Bagnoli92}. Further work in this direction is in progress.

Finally, the PST phase diagram is shown in  Fig.~\ref{fig:PST}, where 
$p^*$ is plotted versus $\eps$. The behavior of $p^*$ is compared with 
that of the damage spreading velocity $V_F$ (Ref~\cite{mappets}),
properly re-scaled.
The consistency of the two phase diagrams, even in the fuzzy
region, suggests that the indicators, $V_F$ and $p^*$, characterize
different aspects of the same phenomenon.
The strict correlation between $V_F$ and $p^*$ is not surprising if one considers
the changing rate of the density of non synchronized sites,
$n(t) = N(t)/L$, in the pinching synchronization mechanism.
For large time and systems and in a mean-field description of the process,
we can write for $n(t)$ an equation similar to that one employed in
contact processes and DP theory \cite{Janns},
\begin{equation}
\dot{n} = 2 V_F n (1-n) - p n\;.  
\label{eq:ndot}
\end{equation}
The first term of the r.h.s. represents the active-site production due to the
linear spreading of desynchronization regions,
which occurs with a velocity $V_F$ (see Eq.~\ref{eq:damage})). 
The second term accounts for the destruction of active sites by the pinching
mechanism.
The process becomes critical when the linear contributions, 
$2 V_F\;n$ and $p\,n$ balance, thus we obtain $p^* \sim V_F$.
The proportionality holds even outside of mean-field approach, i.e., when 
other terms, such as, higher powers of $n$, diffusion and multiplicative 
noise, are included in Eq.~(\ref{eq:ndot}), provided one considers the
renormalized parameters.
The conclusion that $p^*$ is a nondecreasing function of $V_F$ is however 
intuitive, because the higher $V_F$, the grater is the desynchronization rate,
so the pinching probability has to be large to ensure the synchronization.   

\section{Conclusions}
In summary, we have applied the pinching synchronization method
to systems showing stable chaos. As for cellular automata,
we have found that even in this continuous case the 
pinching synchronization transition (PST) is well defined, and that
this transition belongs to the DP universality class.
Our results show that the stable chaos is indeed equivalent to cellular
automata ``chaoticity'' and definitively different from transient chaos. 
The PST phase diagram is consistent with that reported in
Ref.~\cite{mappets} for damage spreading velocities,
including the fuzzy region.

\section*{Acknowledgements}
We thank A.~Politi, R.~Livi, A.~Vespignani and A.~Flammini
for fruitful discussions.
We also thank all the components of the D.O.C.S. research
group of Firenze (http://www.docs.unifi.it) for
stimulating discussions and interactions.


\newpage
\section*{Figure captions}
\renewcommand{\labelenumi}{\bf Figure~\theenumi:}
\renewcommand{\theenumi}{\arabic{enumi}}

\begin{enumerate}
\setlength{\itemsep}{.5cm}

\item {Plot of the map of Eq.~(\ref{eq:stepmap}) for different values of $\alpha$: 
$\alpha=0.02$ (dotted line), $\alpha=0.08$ (dashed line) and $\alpha=1/6$ 
(continuous line).}
\label{fig:stepmap}


\item {Grayscale representation of space-time evolution of 
model of Eq.~(\ref{eq:stepcml}) with $\eps=1/3$ and $\alpha=0.068$. Time runs from top to 
bottom, white (black) color indicates $x_i(t)=0$ ($x_i(t)=1$), while gray 
indicates all other values.}
\label{fig:peda}


\item {Time behaviour of Lyapunov multiplier $\mu(t) = |{\bf z}(t)|/|{\bf z}(t-1)|$ 
for CML of Eq.~(\ref{eq:stepcml}) with $\alpha = 0.052$. The simulation is stopped 
at time $T_r=4455$ when the systems reaches the absorbing SC state.}
\label{fig:multipl}


\item {Histogram of the finite-time Lypunov exponent $\gamma$ of 
model of Eq.~(\ref{eq:stepcml}) computed from a set of $2000$ arbitrary initial 
conditions, with $\alpha=0.052$. 
The size of the system is $L=3000$ sites.  
For each initial condition, the evolution of Eq.~(\ref{eq:stepcml}) and its 
linearization are iterated untill the system approaches a Boolean 
configuration.}
\label{fig:hist}


\item {Plot of the map of Eq.~(\ref{eq:map}).} \hspace{5cm}
\label{fig:figmappet}


\item {Space-time evolution of the map Eq.~\protect\ref{eq:map}, with
$\eps=0.32$. Grayscale from $x_i(t)=0$ (white) to $x_i(t)=1$ (black).}
\label{fig:mappet}


\item {Schematic representation of the pinching synchronization. Vertical
dashed lines indicate the sites of the slave system identified to those 
of the master corresponding to  $r_i=1$  in Eq.~(\ref{eq:slave}).}
\label{fig:pinch}  


\item {Log-log plot of the order parameter $\rho(t,p)$ vs.\ $t$, 
for $p=0.1269,0.1270, 0.1271...$ from top to down, and for $\eps=0.305$. 
The dot-dashed straight line indicates the critical DP scaling ~$t^{-\delta}$ 
with $\delta=0.159$.  The estimated synchronization threshold 
turns to be $p^*=0.1272(1)$.}
\label{fig:rho}  


%
%

\item {effective exponents $\overline{\eta}(t)$ (a), 
$\overline{\delta}(t)$ (b) and $\overline{z}(t)+\overline{\eta}(t)$ (c)
for $\eps=0.3004$ and several values of $p$. 
The average is taken over $50.000$ runs, $a=2$ and $L=2000$.}  
\label{fig:eff}  


\item {Dependence of the synchronization threshold $p^*$ on the
coupling constant $\eps$ (diamonds) compared with the behaviour
of $V_F$ (open circles), the $V_F$-values are properly re-scaled.
In the inset an enlargement of the region around $\eps=0.3$ is shown.
The line is a guide to eye.}
\label{fig:PST}  
\end{enumerate}
\newpage
\pagestyle{empty}

\newcommand{\figref}[1]{Bagnoli and Cecconi,
\emph{Synchronization of non-chaotic ...}\hfill Figure~\ref{#1}}

\includegraphics[width=12cm]{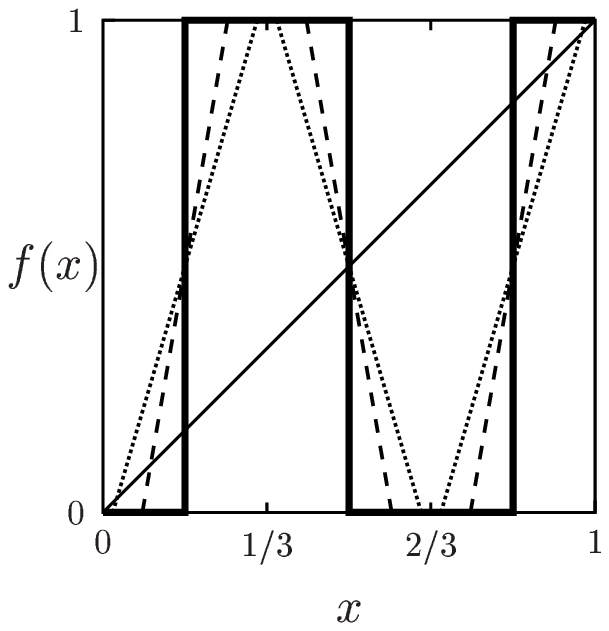}

\vfill
 \figref{fig:stepmap}
\newpage
\includegraphics[width=12cm]{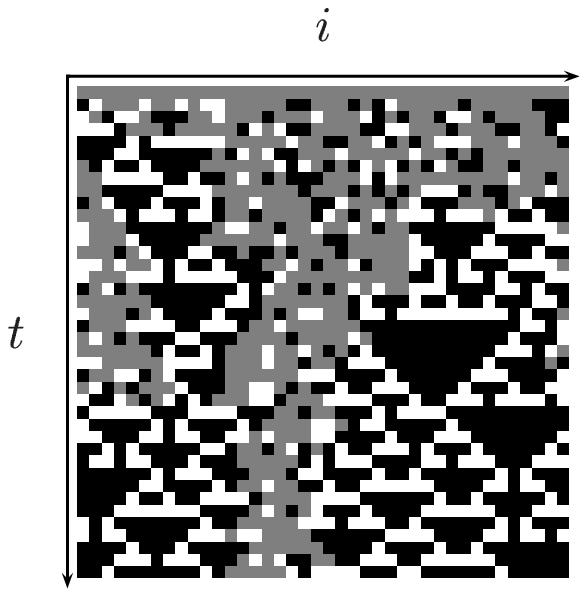}
\vspace{3mm}

\vfill
 \figref{fig:peda}
\newpage
\includegraphics[angle=270,width=12cm]{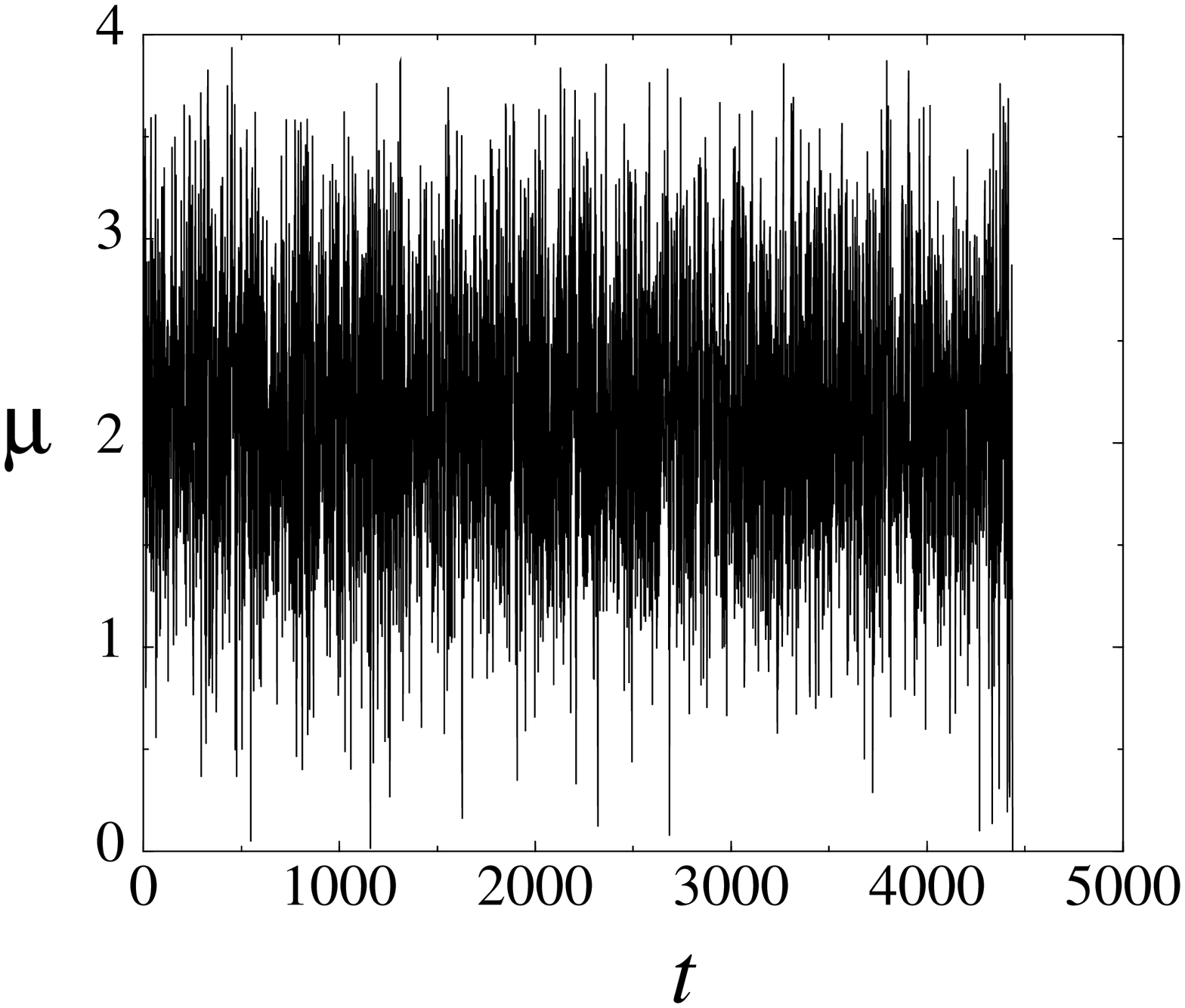}

\vfill
 \figref{fig:multipl}
\newpage
\includegraphics[angle=270,width=12cm]{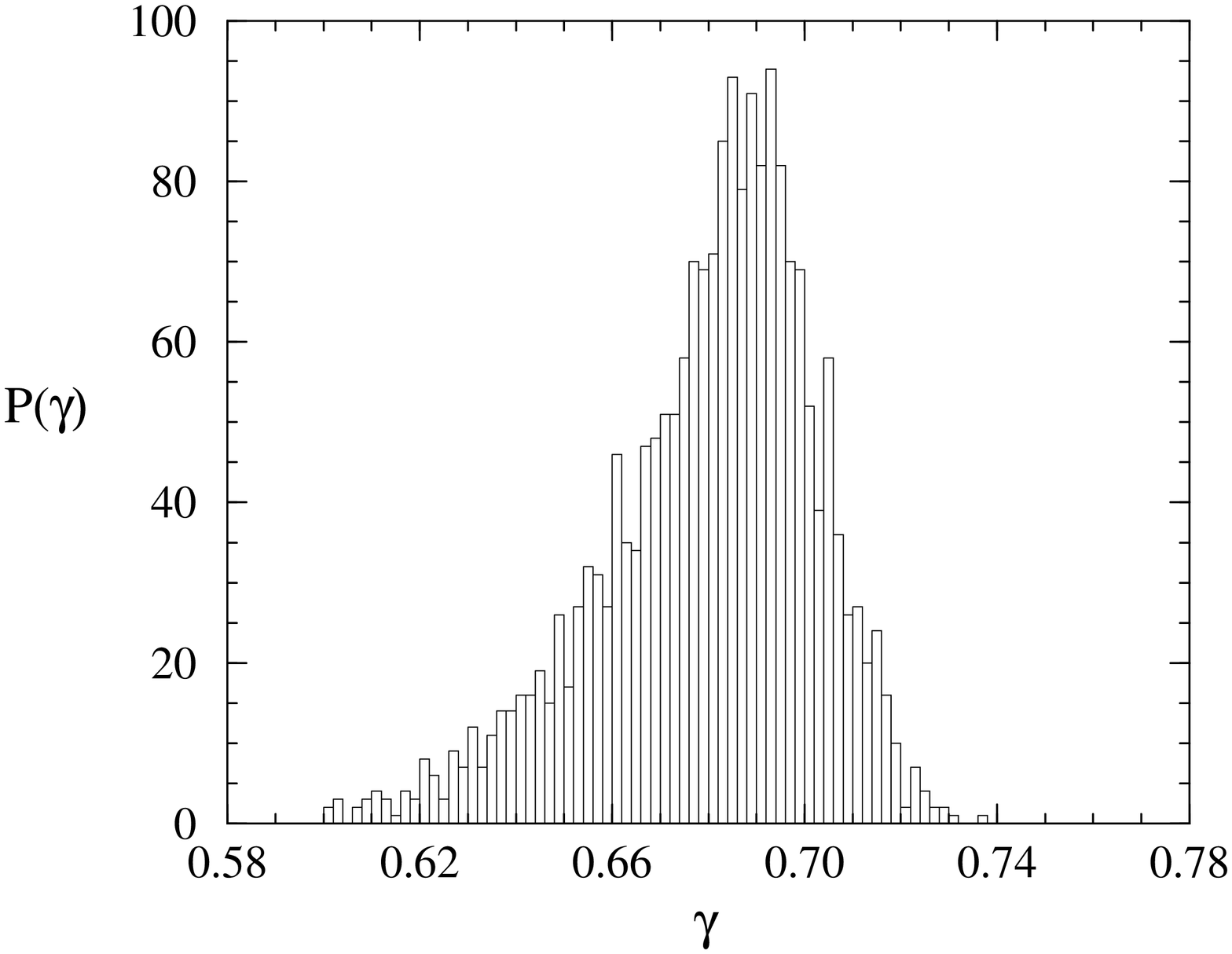}

\vfill
 \figref{fig:hist}
\newpage
\includegraphics[width=12cm]{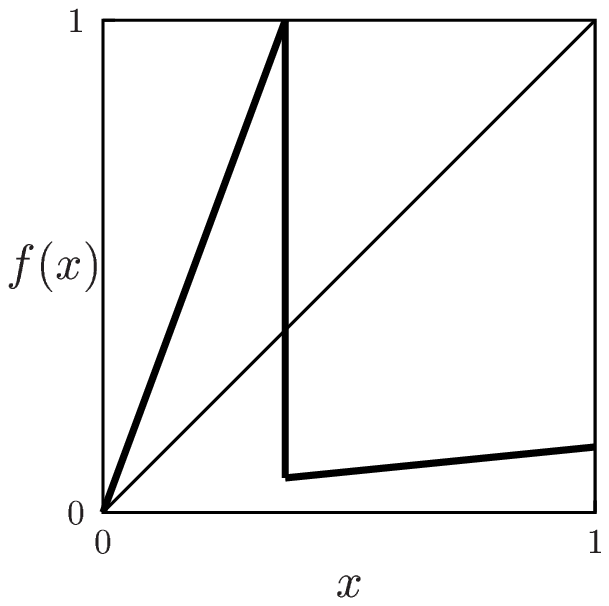}

\vfill
 \figref{fig:figmappet}
\newpage
\includegraphics[width=12cm]{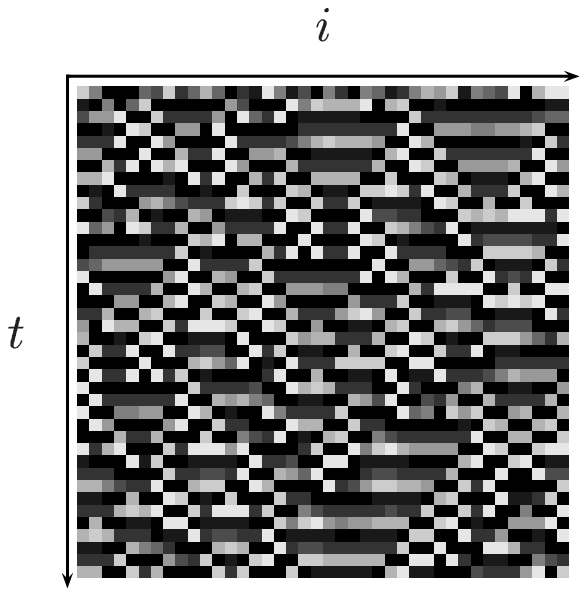}
\vspace{3mm}

\vfill
 \figref{fig:mappet}
\newpage
\bc
\includegraphics[angle=270,width=12cm]{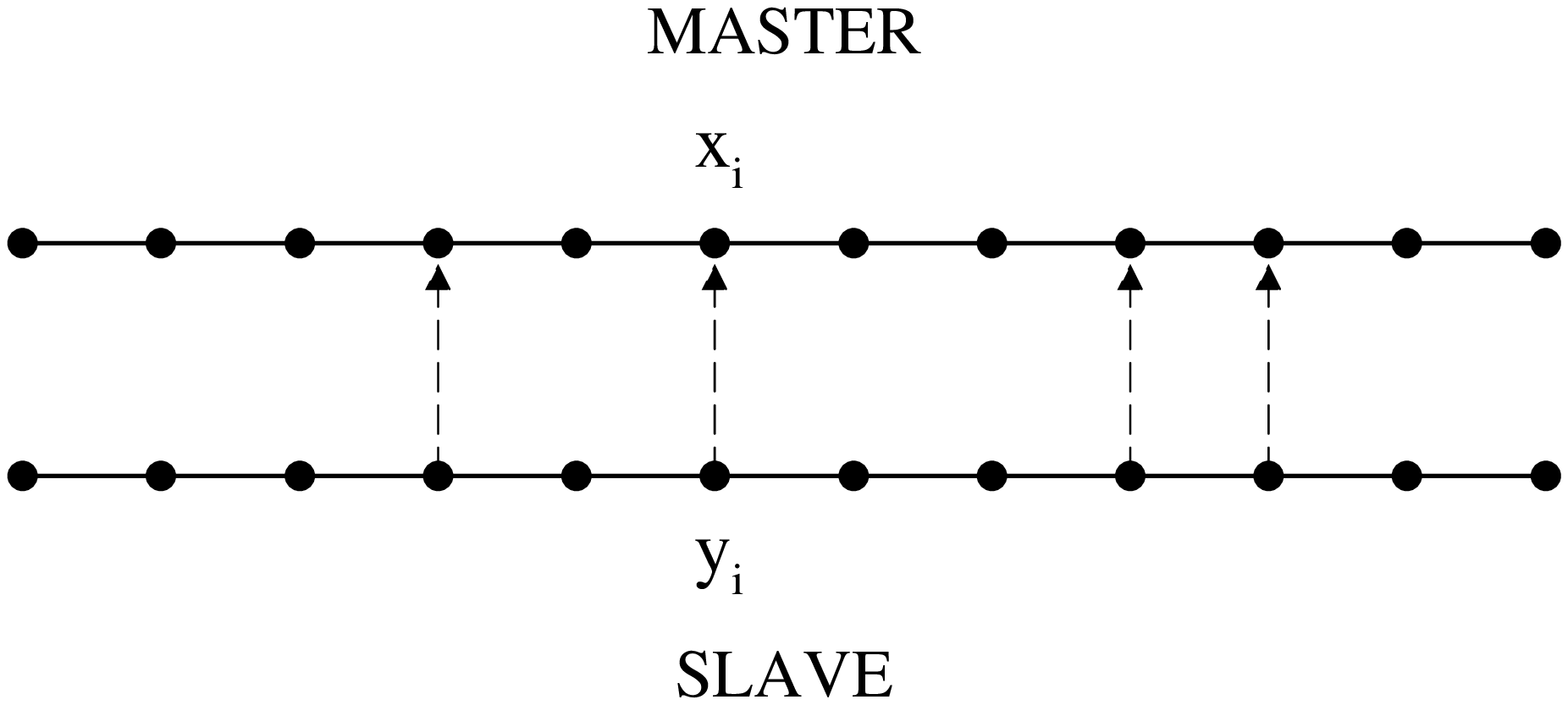}
\ec

\vfill
 \figref{fig:pinch}  
\newpage
\bc
\includegraphics[angle=270,width=12cm]{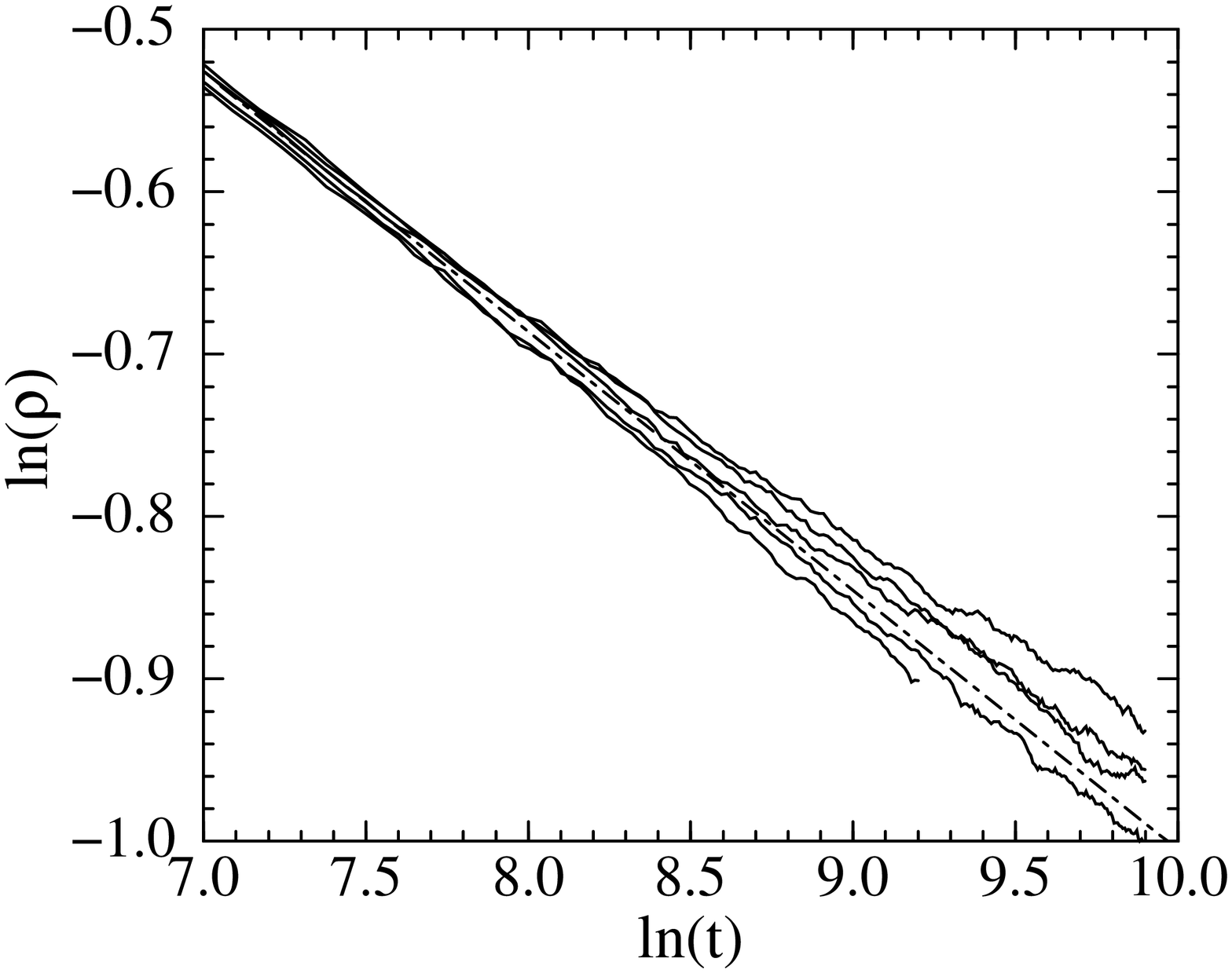}
\ec

\vfill
 \figref{fig:rho}  
\newpage
\hfill \includegraphics[width=0.8\columnwidth]{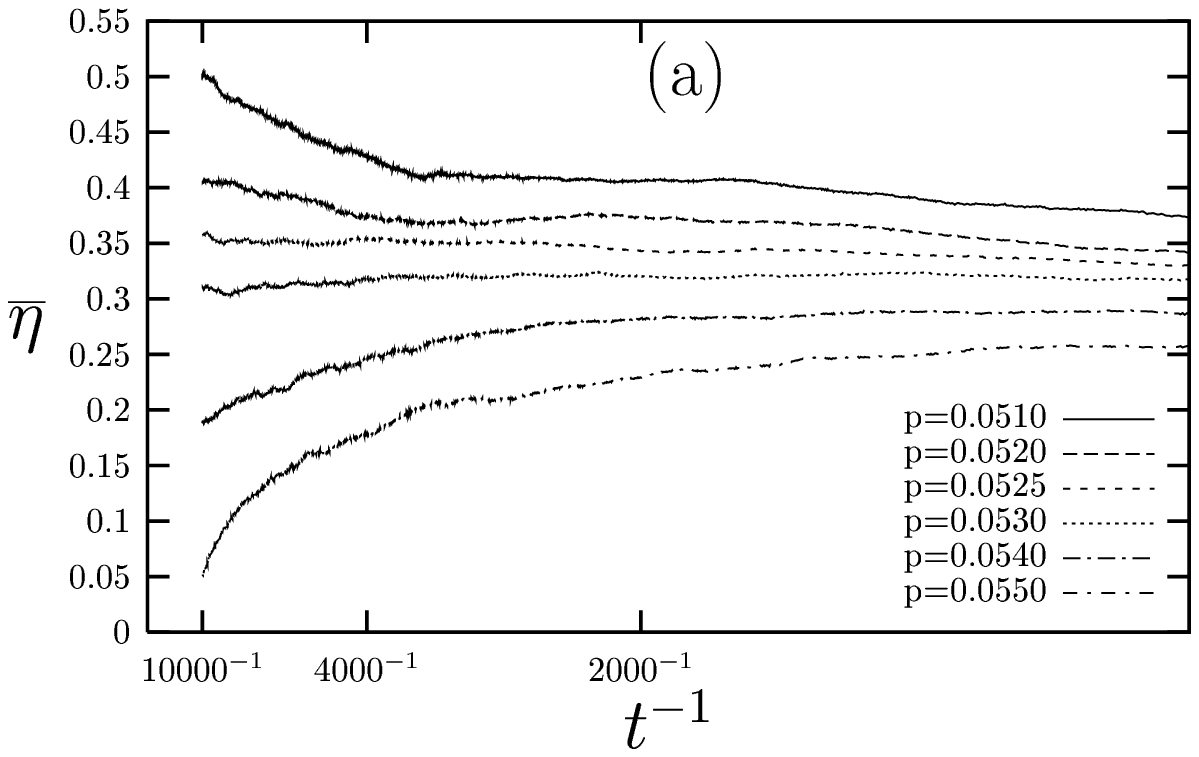}

\hfill \includegraphics[width=0.815\columnwidth]{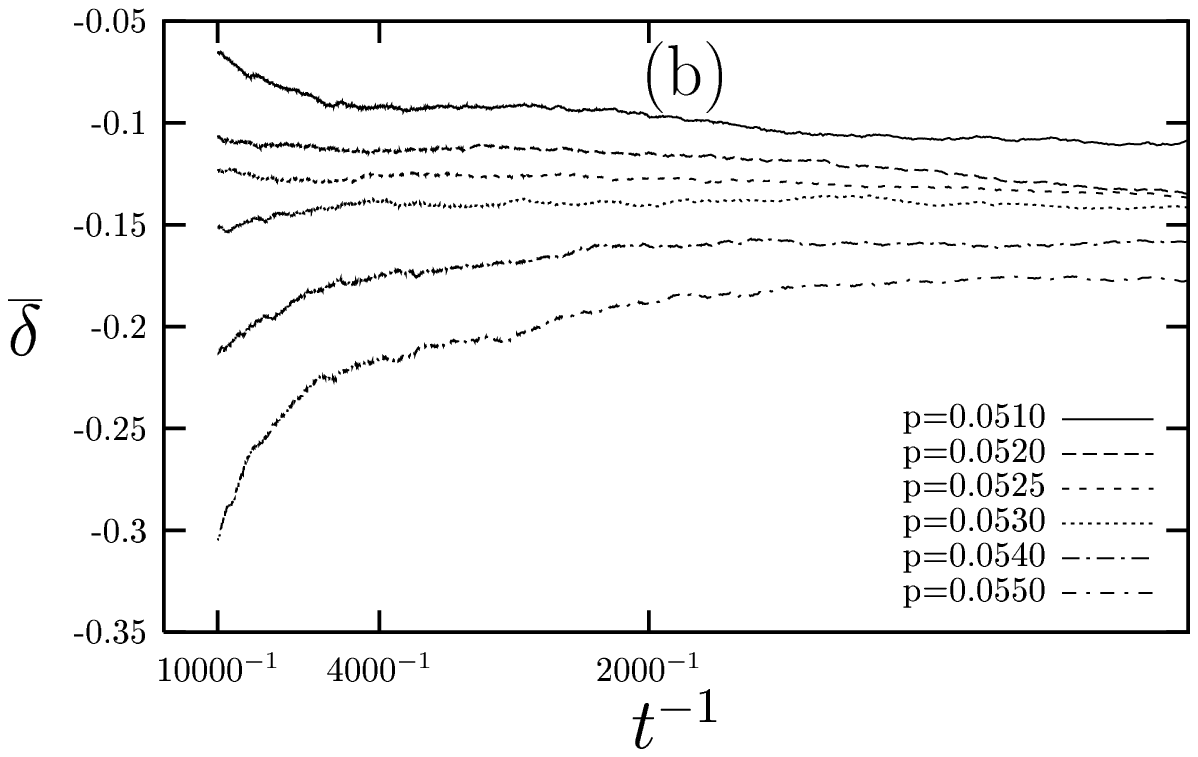}

\hfill \includegraphics[width=0.87\columnwidth]{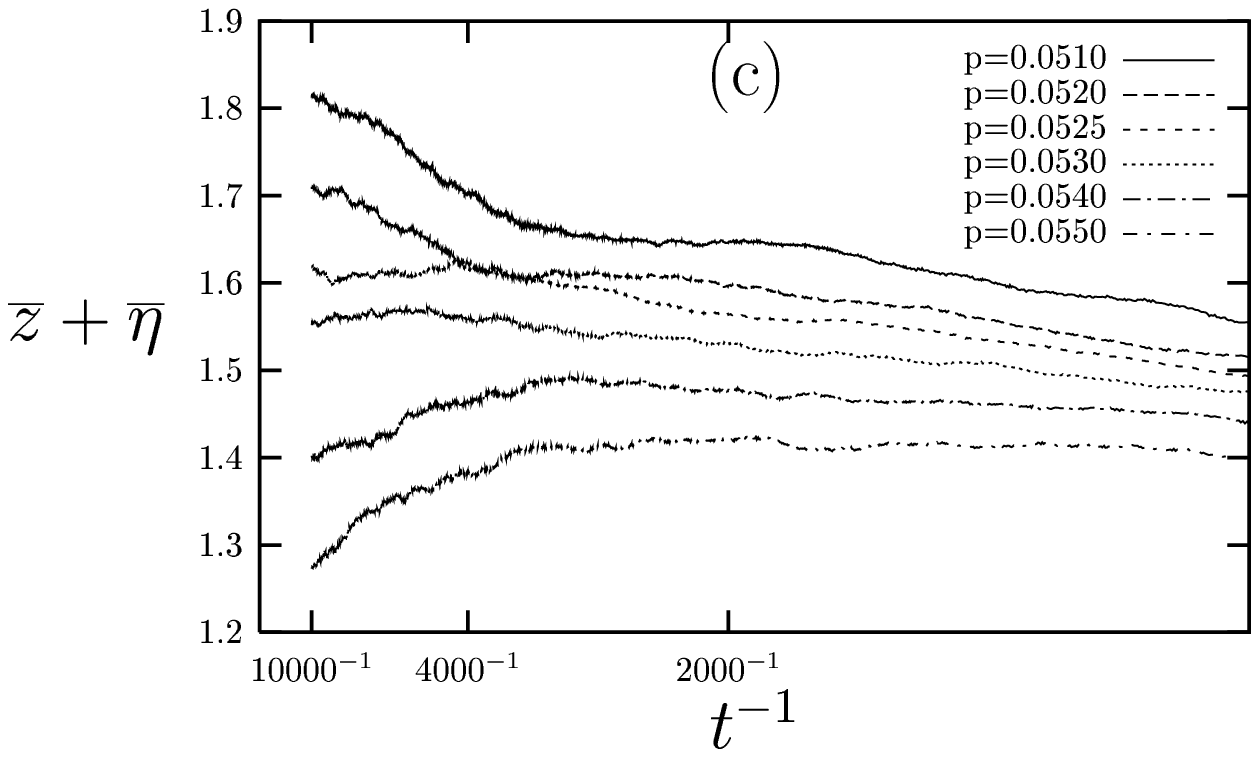}

\vfill
 \figref{fig:eff}  
\newpage
\bc
\includegraphics[angle=270,width=12cm]{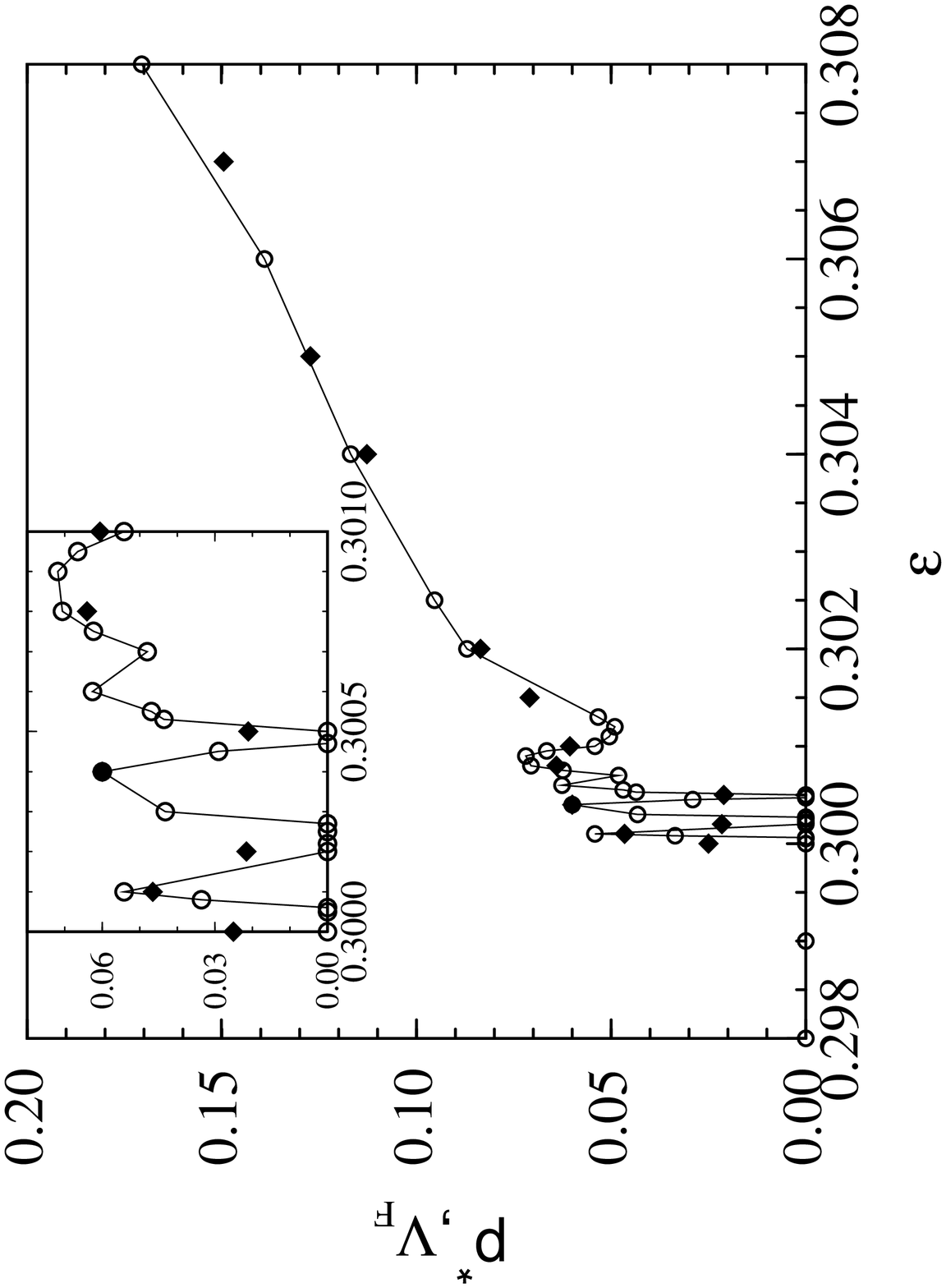}
\ec

\vfill
 \figref{fig:PST}  
\end{document}